\documentclass[aps,prb,twocolumn,superscriptaddress,showpacs]{revtex4-1}

\usepackage{graphicx}
\usepackage{calc}
\usepackage{bm}
\usepackage{color}

\begin{document}

\title{Interlayer current near the edge of an InAs/GaSb double quantum well in proximity with a superconductor}

\author{A.~Kononov}
\affiliation{Institute of Solid State Physics RAS, 142432 Chernogolovka, Russia}
\author{S.V.~Egorov}
\affiliation{Institute of Solid State Physics RAS, 142432 Chernogolovka, Russia}
\author{N.~Titova}
\affiliation{Moscow State Pedagogical University, Moscow 119991, Russia}
\author{B.R.~Semyagin}
\affiliation{Institute of Semiconductor Physics, Novosibirsk 630090, Russia}
\author{V.V.~Preobrazhenskii}
\affiliation{Institute of Semiconductor Physics, Novosibirsk 630090, Russia}
\author{M.A.~Putyato}
\affiliation{Institute of Semiconductor Physics, Novosibirsk 630090, Russia}
\author{E.A.~Emelyanov}
\affiliation{Institute of Semiconductor Physics, Novosibirsk 630090, Russia}
\author{E.V.~Deviatov}
\affiliation{Institute of Solid State Physics RAS, 142432 Chernogolovka, Russia}

\date{\today}

\begin{abstract}
We investigate  charge transport through the junction between a niobium superconductor and the edge of a two-dimensional electron-hole bilayer, realized in an InAs/GaSb double quantum well. For the transparent interface with a superconductor, we demonstrate that the junction resistance is determined by the interlayer charge transfer near the interface. From an analysis of experimental $I-V$ curves we conclude that  the proximity induced superconductivity efficiently couples electron and hole layers at low currents. The critical current demonstrates periodic dependence on the in-plane magnetic field, while it is monotonous for the field which is normal to the bilayer plane.
\end{abstract}

\pacs{73.40.Qv  71.30.+h}

\maketitle

\section{Introduction}

Recent interest to an InAs/GaSb two-dimensional (2D) bilayer system is mostly connected with the problem of a topological insulator~\cite{zhang1,kane,zhang2}. Bulk spectrum with band inversion is realized for the 12~nm thick InAs (electrons) and 10~nm thick GaSb (holes) layers~\cite{gasb3}. Spectrum hybridization~\cite{dqw} is expected at equal carriers' concentrations, so the edge transport is dominant. Supposed to be topological, this one-dimensional edge transport is a subject of continuous attention~\cite{gasb3,gasb1,gasb2,gasb4,gasb5,gasb6}. In  proximity with a superconductor, it is regarded~\cite{su-gasb1,su-gasb2,su-gasb3} to be suitable for Majorana fermion investigations~\cite{reviews}.

On the other hand, interlayer effects are of primary importance in different bilayer systems: Coulomb correlations were shown to be responsible, e.g., for the fractional quantum Hall effect at filling factor $\nu=1/2$~\cite{chak}, and the many-body quantum Hall plateau at $\nu=1$ \cite{murphy}, while the interlayer transport creates a broken-symmetry state~\cite{manoh} and a canted antiferromagnetic state~\cite{vadik}. Recently, serious interest is attracted by the bilayer exciton correlated state~\cite{bilayer_exp1,bilayer_exp2,bilayer_exp3}. 
A  four-particle Andreev process has been predicted~\cite{bilayer_theor} at the interface between a superconductor and a bilayer exciton structure, also in the topological regime~\cite{golubov}. Thus, it seems to be important to study interlayer effects also for a recently popular InAs/GaSb bilayer in proximity with a superconductor.
   
\begin{figure}
\includegraphics[width=0.8\columnwidth]{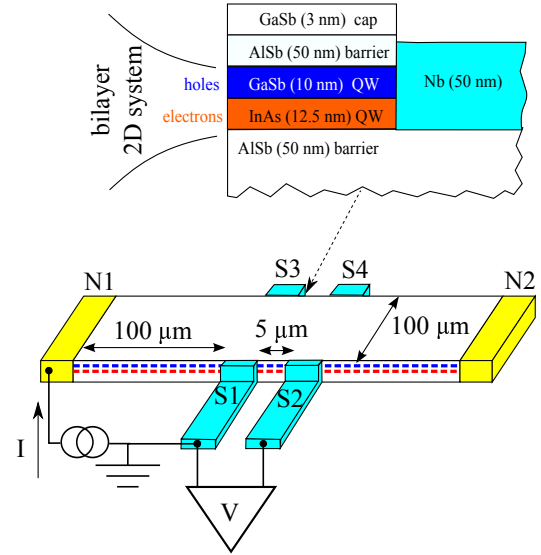}
\caption{(Color online) Sketch of the sample with four niobium contacts to the edge of an InAs/GaSb 2D bilayer  (not to scale). Side NS junctions are fabricated  by sputtering of a thick Nb film (gray) over the mesa step. This film is mostly connected to the bottom (electron) InAs layer (see the text for details). Charge transport is investigated in a four-point technique: one superconducting electrode (S1)  is grounded; a current is fed through the normal Ohmic contact N1; we measure a voltage drop between two superconducting electrodes S1 and S2.  
}
\label{sample}
\end{figure}

Here, we investigate  charge transport through the junction between a niobium superconductor and the edge of a two-dimensional electron-hole bilayer, realized in an InAs/GaSb double quantum well. For the transparent interface with a superconductor, we demonstrate that the junction resistance is determined by the interlayer charge transfer near the interface. From an analysis of experimental $I-V$ curves we conclude that  the proximity induced superconductivity efficiently couples electron and hole layers at low currents. The critical current demonstrates periodic dependence on the in-plane magnetic field, while it is monotonous for the field which is normal to the bilayer plane.

\section{Samples and technique}

Our samples are grown by solid source molecular beam epitaxy on semi-insulating GaAs (100)  substrates. The bilayer is composed of two,  12.5-nm thick InAs and 10-nm thick GaSb, quantum wells (for electrons and holes, respectively), sandwiched between two 50-nm thick AlSb barriers. Details on the growth parameters can be found elsewhere~\cite{growth}. As obtained from standard magnetoresistance measurements, the 2D system is characterized by bulk hole-type conductivity. The mobility at 4K is about $2 \cdot 10^{4}  $cm$^{2}$/Vs  and the carrier density is   $2 \cdot 10^{12}  $cm$^{-2}$.

A sample sketch is presented in Fig.~\ref{sample}. The 100~$\mu$m wide mesa is formed by wet chemical etching. To realize transparent  interfaces with a superconductor, etching  is stopped just after the bottom InAs quantum well (80 nm mesa step samples), or even before it (60 nm  mesa step samples). Ohmic contacts are made by thermal evaporation of 100~nm Au film. They are characterized by a constant, bias-independent $\approx 1 k\Omega$ resistance. 

\begin{figure}
\includegraphics[width=\columnwidth]{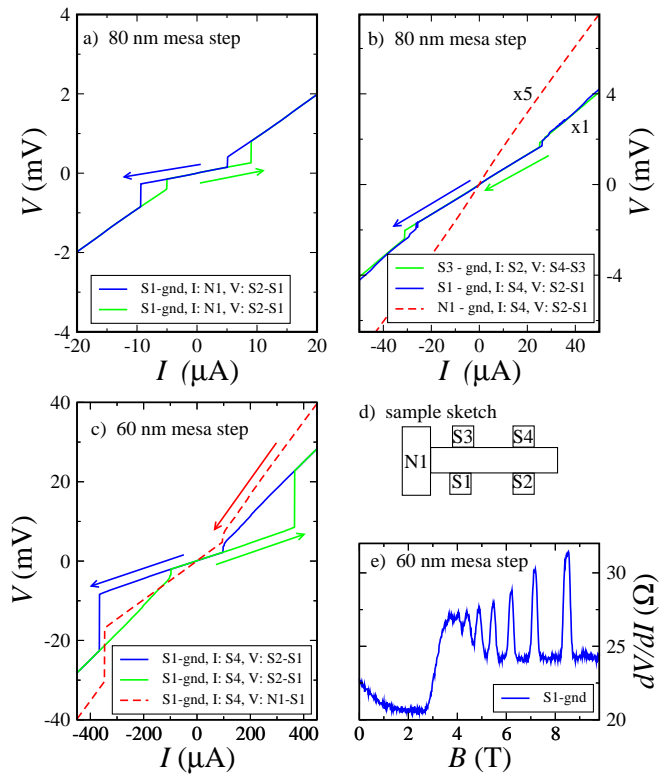}
\caption{(Color online) (a-c) Examples of  $I-V$ characteristics in different experimental configurations  in zero magnetic field. 
	(d) Sample sketch to follow experimental configurations in (a-c).
	(a) The electrical connections are as depicted in Fig.~\protect\ref{sample}. $I-V$s  demonstrate two sharp jumps, which are subjected to hysteresis with the sweep direction (blue and green curves). The voltage drop is not zero between the jumps.   
	(b) $I-V$s coincide well for two symmetric probe connections (green and blue curves). The $I-V$ curve is linear with $\approx 1 k\Omega$ resistance (red dash, please note $\times 5$ coefficient), if the normal (N1) contact is grounded instead of the niobium one. 
	(c) Similar $I-V$ characteristics with hysteresis (green and blue) for the 60~nm mesa step sample.  The $I-V$ still demonstrates qualitatively the same  behavior, if the voltage is taken from the normal contact N1  (red dashed curve).   
	(e) Zero-current differential resistance as a function of normal magnetic field. Magnetoresistance oscillations appear above 3.5~T, just after the superconducting critical field $B_c$=2.5~T. 
	All the curves are obtained at  low temperature  $T=30$~mK$<<T_c$. 
} 
\label{IV}
\end{figure}

Since the edge effects are of primary importance in InAs/GaSb bilayers~\cite{gasb3,gasb1,gasb2,gasb4,gasb5,gasb6,su-gasb1,su-gasb2,su-gasb3}, we fabricate {\em side} superconductor-normal (NS) junctions~\cite{nbhgte,nbsemi} by sputtering  50~nm thick  Nb or NbN film over the  mesa step, see Fig.~\ref{sample}. The surface is mildly cleaned by Ar plasma. To avoid mobility  degradation, the sample is kept at room temperature during the sputtering.  The 10~$\mu$m wide electrodes are formed by lift-off technique. This produces NS junctions with low, below $100 \Omega$, resistance. The junctions are not sensitive to low lithography misalignments, because the upper AlSb layer is an insulator at low temperatures. However, the side Nb contact is mostly connected to the bottom (electron) InAs layer. This is obvious for the 60~nm mesa step sample, where this layer is not removed. It is also the case for the photolithographically fabricated contacts to the 80~nm mesa step  sample, because the common developer etches selectively the GaSb layer~\cite{gasb-etching}.

We study current through the Nb-InAs/GaSb junction in a four-point technique. An example of electrical connections is presented in Fig.~\ref{sample}: one superconducting electrode (S1)  is grounded; a current $I$ is fed through the normal Ohmic contact N1; we measure a voltage drop $V$ between another superconducting electrode S2 and the ground lead (S1). In this connection scheme, a wire to the grounded superconducting S1 contact is excluded, which is necessary for low-impedance NS junctions.

To obtain $I-V$ characteristics, which are presented in  Fig.~\ref{IV},  we sweep the dc current $I$ and measure the voltage drop $V$ in a mV range. To accurately determine critical current values $I_c$, presented in  Fig.~\ref{Ic_BT},  we simultaneously measure $dV/dI(V)$ characteristics: the dc current is additionally modulated by a low ac component (10~nA, 1~kHz), the ac  component of $V$ ($\sim dV/dI$) is detected by a lock-in amplifier. We have checked, that the lock-in signal is independent of the modulation frequency in the 600~Hz -- 3000~Hz range, which is defined by applied ac filters.

To extract features specific to the InAs/GaSb bilayer, the measurements are performed at low temperature of 30~mK. Because of similar critical temperature for Nb and NbN films ($T_c=$9~K and 11~K, respectively, both $T_c>>30$~mK), the  samples demonstrate even quantitatively similar  behavior. However, Nb contacts are preferable, since the much lower niobium critical field $B_c\approx 2.5$~T allows to distinguish between  the bulk and interface effects, see Fig.~\ref{IV} (e).  We will concentrate on these niobium samples below.

\section{Experimental results}

Fig.~\ref{IV} (a-c) presents examples of the $I-V$ characteristics in zero magnetic field for the 80~nm (a,b) and 60~nm (c) mesa step samples. 

The curves in Fig.~\ref{IV} (a) are obtained in the electrical connection scheme, which is depicted in Fig.~\protect\ref{sample}. They  demonstrate  Josephson-like behavior~\cite{tinkham}: two sharp jumps appear at $I_c \approx \pm 10\mu$A, the exact jumps' positions are subjected to hysteresis with the sweep direction (blue and green curves). However, unlike the standard  Josephson $I-V$ behavior~\cite{tinkham}, the voltage drop is not zero between the jumps.  

In our four-point connection scheme, a wire to the grounded superconducting Nb contact is excluded, see Fig.~\protect\ref{sample}. Thus, it can not be responsible for  the non-zero $I-V$ slope between the jumps in Fig.~\ref{IV} (a). On the other hand, the slope corresponds to  $\approx 25 \Omega$ resistance, which  is comparable with the sheet 2D bilayer resistance at present concentration and mobility. 

\begin{figure}
\includegraphics[width=\columnwidth]{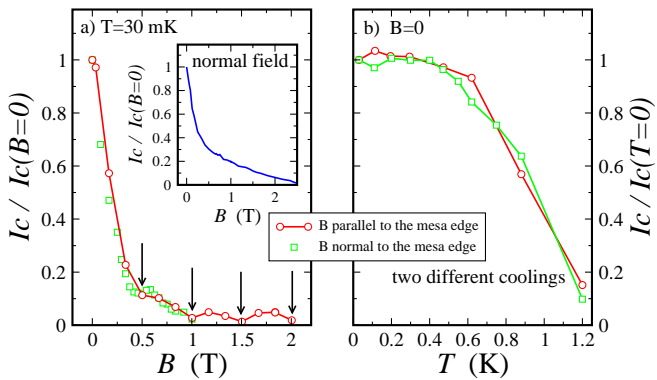}
\caption{(Color online) Critical (jump) current $I_c$ suppression by a magnetic field (a) or temperature (b) for the 80~nm mesa step sample. Because of $I-V$ hysteresis, $I_c$ is defined  as a half-sum of the positive and negative jump positions.  (a) If the magnetic field is within the 2D plane, oscillations in $I_c(B)$ are observed with 0.5~T period for both field orientations with respect to the mesa edge  (normal, green squares,  or parallel, red circles). Inset demonstrates monotonous $I_c$ suppression if the magnetic field is normal to the 2D plane. The curves are obtained at  low temperature $T=30$~mK$<<T_c=9$~K. (b) $I_c$ weakly depends on temperature at low $T<0.5$~K, but sharply falls to zero at higher 0.5~K$<T<$1.2~K. The curves are obtained in zero magnetic field for two sample coolings, which correspond to the experimental configurations depicted in (a). 
} 
\label{Ic_BT}
\end{figure}

Fig.~\ref{IV} (b) demonstrates that the experimental $I-V$s are indeed sensitive to the current distribution within the 2D plane. Two (green and blue) curves are obtained in two symmetric contact configurations, when the current and voltage probes are placed at the opposite sample edges. The curves coincide well, because of symmetric current distribution, but differ strongly from ones in Fig.~\ref{IV} (a):  the jumps' positions are placed at $\approx \pm 30\mu$A and the slope between them corresponds to $\approx 60 \Omega$.

The non-zero voltage at low currents in Fig.~\ref{IV} is inconsistent with the (edge or bulk) Josephson supercurrent between the superconducting potential contacts. Instead, it seems that the jumps on the experimental $I-V$s originate from a single (grounded) Nb-InAs/GaSb junction,  which is connected {\em in-series} with a part of the 2D bilayer system. This behavior is still induced by proximity with a superconductor: if the normal (N1) contact is grounded instead of the niobium one, $I-V$ is of  linear Ohmic behavior, see the red dashed line in Fig.~\ref{IV} (b).  We can expect even stronger proximity effect for the sample with lower (60~nm) mesa step, because the bottom InAs layer is efficiently contacted to the side Nb electrode in this case. In the experiment, the $I-V$ jumps are situated at approximately 10 times higher current, see Fig.~\ref{IV} (c). 

To our surprise, the $I-V$ curve demonstrates the same behavior  if the voltage $V$ is taken from the normal contact N1, which is 100~$\mu$m separated from the grounded superconducting  contact S1, see the red dashed curve in Fig.~\ref{IV} (c).  Thus, the jumps on the experimental $I-V$s reflect transport properties of a single (grounded) Nb-InAs/GaSb junction.

Since the $I-V$ behavior in Fig.~\ref{IV} is induced by  superconductivity, it can be suppressed by a magnetic field or temperature. 

The $I-V$ slope at zero current demonstrates strong increase near the niobium superconducting critical field $B_c$=2.5~T, as demonstrated in Fig.~\ref{IV} (e) for the 60~nm mesa step sample in normal magnetic field. In higher magnetic fields, well-developed Shubnikov-de-Haas magnetoresistance oscillations appear. The latter is a fingerprint of a 2D conducting system, so the $I-V$ curves in Fig.~\ref{IV} reflect charge transport through the InAs/GaSb bilayer to the side superconducting Nb contact.

Suppression of the critical (jump) current $I_c(B)$ is sensitive to the magnetic field orientation, see Fig.~\ref{Ic_BT} (a). The oscillations in $I_c(B)$ with equal $\Delta B=0.5$~T period are observed if the magnetic field is oriented within the 2D plane, either normal (green squares) or parallel (red circles) to the mesa edge. It is worth to mention, that the oscillatory pattern in Fig.~\ref{Ic_BT} (a) even qualitatively differs from the Fraunhofer one~\cite{su-gasb1}, which is another argument against the Josephson effect.  In contrast, $I_c(B)$ is diminishing slower, without any sign of oscillations, in  the   magnetic field which is normal to the 2D plane, see inset to Fig.~\ref{Ic_BT} (a).  
 
For the 80~nm mesa step sample, $I_c$ weakly depends on temperature at low $T<0.5$~K, but sharply falls to zero at higher 0.5~K$<T<$1.2~K, see Fig.~\ref{Ic_BT} (b).  There is only slow (within 5\%)  $I_c(T)$ dependence below 1.2~K for  the 60~nm mesa step sample. Because of the same niobium superconductor, the  experimental $I-V$s seems to be defined by the proximity-induced gap at the Nb-InAs/GaSb interface.

\section{Discussion}

\begin{figure}
\includegraphics[width=\columnwidth]{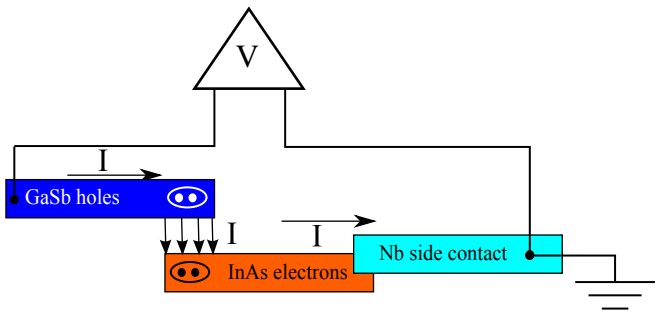}
\caption{(Color online) Interlayer current near the edge of an InAs/GaSb double quantum well in proximity with a superconductor. The bulk of the sample is characterized by hole-type conductivity.  The side Nb contact is mostly connected to the conducting bottom (electron) layer at the edge, so the interlayer charge transport occurs in vicinity of the Nb side contact. At low currents the proximity induced superconductivity efficiently couples electron and hole layers due to the Cooper pair interlayer transfer (see the main text for details). 
}
\label{discussion}
\end{figure}

Above, we present the experimental argumentation that the observed $I-V$ curves are not connected with the Josephson effect. We can also support this conclusion by data known from literature~\cite{su-gasb1,su-gasb2,su-gasb3}. The transition from the bulk to edge Josephson supercurrent was reported in Ref.~\onlinecite{su-gasb1}. The critical current value was about 50~nA for two 400~nm spaced, 4~$\mu$m wide electrodes. This value is in 1000 times smaller than our $I_c$ for electrodes of similar width, separated by 5~$\mu$m distance. Thus, our numerical parameters are inconsistent with the Josephson effect. On the other hand, sharp $I-V$ jumps were reported at high $I_c\approx 50 \mu$A in Ref.~\onlinecite{su-gasb2} for  a single Ta-InAs/GaSb junction, the differential resistance was finite between the jumps. This behavior was out of focus of Ref.~\onlinecite{su-gasb2}, because $I_c$ was identified as a critical current in the superconducting Ta lead. We wish to mention here, that  $I_c$ depends on the carrier concentration in the InAs/GaSb bilayer in Fig.~1 of Ref.~\onlinecite{su-gasb2}, so it more likely should be connected with transport within the InAs/GaSb bilayer.  

We show below from our experimental dependencies, that the observed in Fig.~\ref{IV}  $I-V$ curves reflect the interlayer current near the edge of an InAs/GaSb double quantum well in proximity with a superconductor. 

The bulk of the sample is characterized by hole-type conductivity. By approaching the edge, the hole concentration is gradually  diminishing, because holes screen the edge potential~\cite{shklovskii,image02}. In other words, the edge potential is equivalent to the local gate~\cite{shklovskii,image02}, so both electron and hole layers are conducting near the edge. The side Nb contact is mostly connected to the bottom (electron) layer, see Fig.~\ref{discussion}. This is obvious for the 60~nm mesa step sample, but it is also the case for the photolithographically fabricated contacts to the 80~nm mesa step  sample, because the common developer etches selectively the GaSb layer~\cite{gasb-etching}.

Thus, the experimental $I-V$ curve reflects not only in-series connected resistances of the 2D hole layer and the Nb-InAs interface, but also the interlayer charge transfer  in the vicinity of the Nb contact, see Fig.~\ref{discussion}. The latter term is dominant for the transparent interface and high mobility holes. The  proximity induced superconductivity can efficiently couple electron and hole layers 
 due to the Cooper pair transfer. Thus, at low currents the experimental $I-V$ curve reflects the in-plane resistance of the 2D hole gas. The sharp jumps at high $I_c$ are defined by the destruction of the coherent interlayer Cooper pair transport. This situation is different from the proposals of Refs.~\cite{bilayer_theor}, where interlayer coupling occurs due to the exciton condensate formation.

The proposed interpretation explains why the experimental $I-V$s are sensitive to the current distribution within the 2D plane in Fig.~\ref{IV} (a) and (b). It is also strongly confirmed by the $I_c$ dependencies on the mesa step hight and temperature. 

(i) For the 60~nm mesa step sample, the bottom InAs layer is efficiently contacted to the side Nb electrode, which results in the maximum proximity effect. The deeper the mesa etching, the lower the area of the Nb-InAs contact, and, therefore~\cite{tinkham,BTK}, lower the induced superconductivity. It results in lower $I_c$ for the 80~nm mesa step sample in Fig.~\ref{IV}.

(ii) The induced superconducting gap $\Delta$ can be estimated for the interlayer transport region from the temperature dependence of $I_c$. 	 From Fig.~\ref{Ic_BT} (b),  $\Delta$ is about 1.3~K for the 80~nm mesa step sample.  Since there is only slow (within 5\%)  $I_c(T)$ dependence below 1.2~K for  the 60~nm mesa step sample, $\Delta$ is close to the bulk value 9~K in this case. 

The periodic suppression of $I_c$ by magnetic field is more complicated. The oscillations in $I_c(B)$ suppose a magnetic flux $\Phi_0$ penetration through the closed loop. It seems to be possible in the geometry of Fig.~\ref{discussion} if the magnetic field is oriented along the mesa edge. The field induces a phase difference between the electron and hole layers, which results in the $I_c(B)$ oscillations. The experimentally observed period  $\Delta B=0.5$~T in Fig.~\ref{Ic_BT} (a)   corresponds ($S\Delta B \sim \Phi_0$) to the effective area $S\approx 10^{-10}\mbox{cm}^{-2}$. If we assume the effective layers' spacing $d$ as 10~nm, the lateral dimension of the edge region in Fig.~\ref{discussion} can be estimated as $S/d\approx 1\mu$m. This value is consistent with the proximity-induced superconductivity, because of the niobium  coherence length $\xi^0_{Nb}=\hbar v_F/\Delta_{Nb} \approx 1 \mu$m. The region of coherent interlayer current is of  the axial symmetry near a side Nb contact, so the $I_c(B)$ oscillations could be expected for any magnetic field orientation within the 2D plane.  However, the interlayer phase difference does not occur in normal magnetic field, so we observe monotonous suppression of $I_c$ in the inset to Fig.~\ref{Ic_BT} (a).

\section{Conclusion}

As a conclusion, we investigate  charge transport through the junction between a niobium superconductor and the edge of a two-dimensional electron-hole bilayer, realized in an InAs/GaSb double quantum well. For the transparent interface with a superconductor, we demonstrate that the junction resistance is determined by the interlayer charge transfer near the interface. From an analysis of experimental $I-V$ curves we conclude that  the proximity induced superconductivity efficiently couples electron and hole layers at low currents. The critical current demonstrates periodic dependence on the in-plane magnetic field, while it is monotonous for the field which is normal to the bilayer plane. 

\acknowledgments
We wish to thank Ya.~Fominov, V.T.~Dolgopolov, and T.M.~Klapwijk for fruitful discussions.  We gratefully acknowledge financial support by the RFBR (project No.~16-02-00405), RAS and the Ministry of Education and Science of the Russian Federation under Contract No. 14.B25.31.0007.


\begin{thebibliography}{99}


\bibitem{zhang1} S. Murakami, N. Nagaosa, S.-C. Zhang, Phys. Rev. Lett. 93, 156804 (2004).
\bibitem{kane} C. L. Kane, E. J. Mele, Phys. Rev. Lett. 95, 146802 (2005).
\bibitem{zhang2} B. A. Bernevig, S.-C. Zhang, Phys. Rev. Lett. 96, 106802 (2006).

\bibitem{gasb3} K. Suzuki, Y. Harada, K. Onomitsu, and K. Muraki, Phys.  Rev. B 87, 235311 (2013).

\bibitem{dqw} V. T. Dolgopolov, A. A. Shashkin, E. V. Deviatov,
 F. Hastreiter, M. Hartung, A. Wixforth, K. L. Campman, and A. C. Gossard
Phys. Rev. B 59, 13235 (1999)

\bibitem{gasb1} C. Liu, T. L. Hughes, X.-L. Qi, K. Wang, and S.-C. Zhang,  Phys. Rev. Lett. 100, 236601 (2008).
\bibitem{gasb2} I. Knez, R.-R. Du, and G. Sullivan, Phys. Rev. Lett. 107,  136603 (2011).
\bibitem{gasb4} I. Knez, C. T. Rettner, S.-H. Yang, S. S. P. Parkin, L. Du,  R.-R. Du, and G. Sullivan, Phys. Rev. Lett. 112, 026602
 (2014).
\bibitem{gasb5} E. M. Spanton, K. C. Nowack, L. Du, G. Sullivan, R.-R.  Du, and K. A. Moler, Phys. Rev. Lett. 113, 026804 (2014).
\bibitem{gasb6} L. Du, I. Knez, G. Sullivan, and R.-R. Du, Phys. Rev.  Lett. 114, 096802 (2015).


\bibitem{su-gasb1}  V. S. Pribiag,	A. J. A. Beukman,	F. Qu,	M. C. Cassidy,	Ch. Charpentier,	W. Wegscheider	and L. P. Kouwenhoven       Nature Nanotechnology 10, 593 (2015)     doi:10.1038/nnano.2015.86 
\bibitem{su-gasb2} Wenlong Yu, Yuxuan Jiang, Chao Huan, Xunchi Chen, Zhigang Jiang, Samuel D. Hawkins, John F. Klem, and
Wei Pan, Appl. Phys. Lett. 105, 192107 (2014); doi: 10.1063/1.4901965  
\bibitem{su-gasb3} Xiaoyan Shi, Wenlong Yu, Zhigang Jiang, B. Andrei Bernevig, W. Pan, S. D. Hawkins, and J. F. Klem, J. Appl. Phys. 118, 133905 (2015); doi: 10.1063/1.4932644 

\bibitem{reviews} For recent reviews, see C. W. J. Beenakker, Annu. Rev. Con. Mat. Phys. 4, 113 (2013) and J. Alicea, Rep. Prog. Phys. 75, 076501 (2012).

\bibitem{chak} T.~Chakraborty, and P.~Pietilainen, Phys.\ Rev.\ Lett.\ {\bf 59}, 2784 (1987); 
D.~Yoshioka, A.H.~MacDonald, and S.M.~Girvin, Phys.\ Rev.\ B\ {\bf 39}, 1932 (1989); 
J.P.~Eisenstein, G.S.~Boebinger, L.N.~Pfeiffer, K.W.~West, and S.~He, Phys.\ Rev.\ Lett.\ {\bf 68}, 1383 (1992); 
Y.W.~Suen, L.W.~Engel, M.B.~Santos, M.~Shayegan, and D.C.~Tsui, Phys.\ Rev.\ Lett.\ {\bf 68}, 1379 (1992); 
Y.W.~Suen, H.C.~Manoharan, X.~Ying, M.B.~Santos, and M.~Shayegan, Phys.\ Rev.\ Lett.\ {\bf 72}, 3405 (1994);
\bibitem{murphy} S.Q.~Murphy, J.P.~Eisenstein, G.S.~Boebinger, L.N.~Pfeiffer, and K.W.~West, Phys.\ Rev.\ Lett.\ {\bf 72}, 728
(1994); T.S.~Lay, Y.W.~Suen, H.C.~Manoharan, X.~Ying, M.B.~Santos, and M.~Shayegan, Phys.\ Rev.\ B\ {\bf 50}, 17725 (1994).
\bibitem{manoh} H.C.~Manoharan, Y.W.~Suen, T.C.~Lay, M.B.~Santos, and M.~Shayegan, Phys. Rev. Lett., vol. 79, p. 2722 (1997).
\bibitem{vadik} V. S. Khrapai, E. V. Deviatov, A. A. Shashkin,  V. T. Dolgopolov, F. Hastreiter, A. Wixforth, K. L. Campman,
  and A. C. Gossard, Phys. Rev. Lett. {\bf 84}, 725 (2000).

\bibitem{bilayer_exp1} I. B. Spielman, J. P. Eisenstein, L. N. Pfeiffer, and K. W. West, Phys. Rev. Lett. 84, 5808 (2000).
\bibitem{bilayer_exp2} I. B. Spielman, J. P. Eisenstein, L. N. Pfeiffer, and K. W. West, Phys. Rev. Lett. 87, 036803 (2001). 
\bibitem{bilayer_exp3} J. P. Eisenstein  and A. H. MacDonald, Nature (London) 432, 691 (2004) and works cited therein.


\bibitem{bilayer_theor} F. Dolcini, D. Rainis, F. Taddei, M. Polini, R. Fazio, and A. H. MacDonald, Phys. Rev. Lett. 104, 027004 (2010);  S. Peotta, M. Gibertini, F. Dolcini, F. Taddei, M. Polini, L. B. Ioffe, R. Fazio, and A. H. MacDonald, Phys. Rev. B 84, 184528 (2011).
\bibitem{golubov} M. Veldhorst, M. Hoek, M. Snelder, H. Hilgenkamp, A. A. Golubov, and A. Brinkman, Phys. Rev. B 90, 035428 (2014).


\bibitem{growth}   E.A. Emel’yanov, D.F. Feklin, A.V. Vasev, M.A. Putyato, B.R. Semyagin, A.P. Vasilenko, O.P. Pchelyakov, V.V. Preobrazhenskii, Optoelectronics, Instrumentation and Data Processing, 47, 452 (2011).

\bibitem{nbhgte} A. Kononov, S. V. Egorov, N. Titova, Z. D. Kvon, N. N. Mikhailov, S. A. Dvoretsky, E. V. Deviatov, JETP Lett., 101 , 41 (2015).
\bibitem{nbsemi} A. Kononov, S. V. Egorov, Z. D. Kvon, N. N. Mikhailov, S. A. Dvoretsky, and E. V. Deviatov,
 Phys. Rev. B 93, 041303(R) (2016)

\bibitem{gasb-etching} F. Rahman, B. L. Gallagher, M. Behet, and J. De Boeck, Appl. Phys. Lett. 73, 88 (1998); doi: 10.1063/1.121789


\bibitem{tinkham} M. Tinkham, Introduction to Superconductivity (2d ed., McGraw–Hill, New York, 1996).
\bibitem{BTK} G.E. Blonder, M. Tinkham, T.M. Klapwijk, Physical Review B, 25, 4515 (1982).



\bibitem{shklovskii} D. B. Chklovskii, B. I. Shklovskii, and L. I. Glazman, Phys. Rev. B {\bf 46}, 4026 (1992).
\bibitem{image02} E. Ahlswede, J. Weis, K. v. Klitzing, K. Eberl, Physica E,
{\bf 12}, 165 (2002).


\end{thebibliography}
\end{document}